# Optically Induced Polarization and Depolarization of an Electron Spin in a Nitrogen-Vacancy Center of Diamond Nanocrystals

Robert Chapman and Taras Plakhotnik*

School of Mathematics and Physics, The University of Queensland, QLD 4072, Australia

* taras@physics.uq.edu.au

**Abstract**

Pulsed laser excitation causes the luminescence of Nitrogen Vacancy centers in diamond to unexpectedly decrease with increasing pulse energy. This decrease is observed in both the negatively charged and neutral centers and is caused by shortening of the luminescence lifetimes of the centers of both types. In darkness, the luminescence does not show any recovery on a time scale of 10 microseconds but as little as three low-intensity pulses can return the luminescence to its previous, brighter state. An external magnetic field reduces the magnitude of the effect. A possible mechanism for these phenomena based on optical depolarization of an electronic spin is proposed.

## I. INTRODUCTION

Nitrogen-vacancy (NV) centers[1] are defects of the diamond crystal lattice made of a vacancy and an adjacent substitutional nitrogen atom. These centers have recently attracted much interest in many areas of research due to their magneto-sensitivity,[2] photo stability,[3] and chemical inertia.[4] The centers have a large absorption cross-section of $0.95\pm0.25\times10^{-16}$ cm$^{-2}$ at a practically convenient 532-nm wavelength.[5] These unique properties make the NV center a promising candidate for various applications.[6] Nano-magnetometery has been first proposed as a theoretical concept[7] and later demonstrated experimentally.[8, 9] Quantum information processing[10-13] and biomedical applications[14-16] are also hot topics and the provided references give several examples in the area. In recent papers, it has also been demonstrated that nano-diamonds can be used as nano-scale temperature-sensors[17, 18] and electric-field sensors[19] and that the photo stability of NV centers depends on the crystal size and in ultra small (about 5 nm across) diamonds they are subject to luminescence intermittency.[20] A lot of interest exists in fabrication of diamonds with high concentration of NV centers. For example, a high concentration is required for making bright luminescent, 5-nm diamond crystals.[21, 22] Many applications of NV centers are based on unique properties of the electron spin associated with the negatively charged form of NV centers, NV⁻.

## II. SPIN OF NV-CENTERS AND LUMINESCENCE

The total spin of an NV⁻ center in its ground electronic state is 1 and therefore it can have three possible projections, $m = 0$ and $m = \pm 1$ on the axis connecting the nitrogen atom and the vacancy (see Fig. 1). The zero-field splitting between the $m = 0$ and $m = \pm 1$ levels is approximately 2.87 GHz in the electronic ground state. Under normal conditions at room temperature all three spin sublevels are equally populated. The ground triplet state has a strong optical dipole-transition to the electronically excited triplet state which has a similar spin structure and about two times smaller zero-field splitting between the $m = 0$ and $m = \pm 1$ levels.[23] Although absorption of a photon and relaxation to the ground state through photoluminescence does not change the value of $m$ due to the selection rules of the radiative electric-dipole transition, quite unusually, a process initiated by light absorption transfers the spin population from $m = \pm 1$ states to the $m = 0$ state. This process is called optically induced spin polarisation. The accepted model of spin polarization assumes that the states $m = \pm 1$ can relax non radiatively to the ground state via an intermediate singlet electronic state (this state is situated between the two triplet states) and that the singlet state then relaxes predominantly to the $m = 0$ level of the ground state.[24, 25] The described non-radiative path is very inefficient for the $m = 0$ state, causing it to relax primarily via the photoluminescence path. A center polarized in the $m = 0$ state will therefore have a greater photoluminescence intensity than a depolarized center.[2] Although 100% spin-polarization is highly desirable for applications, only about 85% population of the *z*-state has been reported in the literature. It is believed but not confirmed experimentally that the limit of 85% is set by the nonzero probability of the spin-changing optical transitions.[25] A hot topic in the literature is the detailed mechanism of such spin polarization. In particular, the number of singlet states involved and their symmetry has been a subject of scrutiny.[25-30] The general consensus is that the electronic ground term is of $^3A_2$ symmetry ($C_{3v}$ point group) split into a singlet (A symmetry) and a doublet ($E_x$, $E_y$). The doublet is split due to a linear strain in the crystal.[31] This splitting is only about 20 GHz but can be observed at low temperatures. However, electronically excited states are generally not well understood. The unresolved questions are the number of singlet states between the two lowest triplet states, their symmetry and the selection rules for the spin-polarizing intersystem-crossing. The number of the singlet states ranges between one and three in different models. For example, three single states $^1E'$, $^1A$, and $^1E$ (in the order from the highest to the lowest) were recently predicted using *ab initio* many-body perturbation theory.[29] In this work it was suggested that there are two paths for intersystem crossing in NV centers. Both are effective only for $m = \pm 1$ electronically excited states but (depending on the symmetry of the intermediate singlet state) the relaxation path ends at different spin sub-states of the electronic ground state.

In this paper we investigate a new phenomena recently reported[32] which manifests itself as a drop in the luminescence intensity excited by a pulsed laser when the excitation energy density of the excitation pulse increases several times above its saturation value (characteristic for each NV-center). We associate this observation with a new manifestation of the spin polarization in these centers.

It is expected that the luminescence of the NV centers should rise linearly with laser excitation at very low energies, followed by a decrease in the growth rate as the NV in diamond asymptotically approaches its maximum luminescence achieved when a center is excited with probability 1 by each pulse. When the pulse length is much shorter and the delay between the pulses rate is much longer than the relaxation times in the NV center, luminescence saturation can be described by a simple two-parameter equation[5, 32]

$$R = R_\infty \left[1 - \exp\left(-\frac{E}{E_{sat}}\right)\right], \qquad (1)$$

where $R$ is the photon detection rate, $R_\infty$ is its asymptotic value which depends on the quantum yield of the luminescence and the photon detection efficiency of the experimental apparatus, $E$ is the energy density created by a single laser pulse at the location of the NV center, and $E_{sat}$ is a parameter called the saturation energy density which depends only on the absorption cross-section of the NV-center. Saturation of luminescence with a short-pulse laser and Eq. (1) have been successfully used[5] to measure accurately the absorption cross-section of $NV^-$ centers. It was therefore most surprising to find significant deviations from Eq. (1).

**III. EXPERIMENTAL**

The experimental apparatus[5] is based on a wide-field epi-fluorescence microscope. This allows observation of several crystals simultaneously. We focus 532-nm light from a pulsed-output fiber laser (Fianium), through a prefocussing lens and a microscope objective (Nikon, NA 0.9 100X), to form a spot approximately 30 μm in diameter on a quartz slide spin coated with diamond nano-crystals. These crystals have an average size of 30 nm and were purchased from the Academia Sinica production facility. NV centers were produced by irradiation of diamond nano-crystals with $He^+$ ions followed by annealing as described.[33] Luminescence from $NV^-$ centers has been collected by the microscope objective and sent to a detector. Depending on the experiment, for detection we used either a thermoelectrically cooled EMCCD (Andor iXon) or a time-gated Intensified CCD (Stanford Computer Optics 4Picos), which has a sub nanosecond time resolution. Both detectors were used in conjunction with either a spectrometer (Acton SP2300) or a set of filters which transmit emission of negatively charged $NV^-$ in the band 675-700 nm and block both background light and light from the $NV^0$ species, whose luminescence is centered around 630 nm. A magnetic field used in one of the experiments was created by a small permanent magnet. The direction of the field was parallel to the optical axis of the microscope objective and perpendicular to the substrate with the diamond crystals.

**IV. RESULTS**

As has been already mentioned above, the experimental data accurately follows the theory at relatively low pulse energies. This is shown in Fig. 2. Note that the fit made for the 8 lowest pulse energies predicts the position of the next three data points. This fit allows us to characterize each crystal in terms of the number of embedded NV-centers. If all NV-centers were identical, $R_\infty$ would be strictly proportional to the number of NV-centers in a particular crystal. In practice, NV⁻-centers may have different luminescence quantum yields and the collection efficiency of the microscope objective depends on the particular direction of their emission dipole moments. Nevertheless, the distribution of $R_\infty$ obtained by fitting many saturation curves to Eq. (1) has very distinct peaks as shown in Fig. 3. From the histogram we conclude that the average number of NV⁻-centers per crystals in the batch is about 3 and that $R_\infty$ of 0.003 photon/pulse corresponds to 1 negatively charged NV-center, in reasonable agreement with our theoretical estimate of the detection efficiency of our setup.[5] Looking at $R_\infty$ for the crystal whose saturation curve is shown in Figure 2, we see that this crystal is likely to contain 5 NV⁻ centers.

When the pulse energy is approximately 5 times the saturation energy, the experimental points deviate significantly from the theoretical prediction. The emission rate decreases 1.7-fold at 1.1 µJ of pulse energy and drops 1.8-fold at energy of 1.7 µJ. For this crystal, luminescence intensity levels out at approximately half of its maximum value as the pulse energy increases. Note, that the saturation curve was measured by taking half of the measurements going from low-energy to high-energy pulses, and the other half using decreasingly powerful pulses. The fact that the curve remained consistent irrespective of the direction in which the energy changed suggests that there is some mechanism by which the luminescence recovers.

A possible explanation for the drop in the luminescence intensity is light induced conversion of the negatively charged centers to their neutral form. Photoionization of $NV^-$ centers has been experimentally investigated previously.[34, 35] Interestingly, but photo transformation of $NV^0$ to $NV^-$ has also been reported[36] and explained by ionization of an impurity whose electron is then captured by $NV^0$. If excitation with high-intensity pulses causes the $NV^-$ center to ionize, the centers would no longer be visible to our apparatus tuned to detect only $NV^-$ emission, but such ionization would be clearly visible in the emission spectrum. We have measured luminescence spectra for the crystal whose saturation curve is shown in Figure 2. These spectra are shown in Fig. 4. Noting that both spectra are normalized to the same total area, we see that the negatively charged NV center (distinguishable by its 638 nm zero phonon line and a phonon bad with a maximum at about 680 nm) shows a 15% drop in its contribution to the spectrum, while the neutral center (with ~575-nm zero phonon line) shows a 15% increase in its contribution. If the photoionization were not accompanied by any other change in the sample (we will discuss possibilities below), then such a process would be responsible for only 15% decrease in the $NV^-$ signal which is much less than the decrease of the intensity in Fig. 2. In fact, the spectra indicates that emission of $NV^0$ also decreases.

Saturation curves were measured for the neutrally charged NV centers using a filter transmitting light in the band 580-620 nm. It was observed that these centers showed a luminescence decrease similar to $NV^-$. For example, an intensity drop of 1.5 fold at the highest pulse energy was detected for the crystal used to collect the data shown in Figs. 2 and 4.

Note also insignificant change of the zero phonon lines in the spectra shown in Fig. 4. The zero-phonon lines associated with both negatively charged and neutral NV-centers broadens and becomes indistinguishable from the background of the photon wing at temperatures above 500K[17] while only little broadening (less than 30%) has been observed in these experiments. Therefore the average temperature of the crystal should be not more than 100K above room temperature. But in principle it can rise much more for a time much shorter than the luminescence lifetime.

Luminescence intermittency provides another possible explanation for the luminescence drop. While observations of light-induced "blinking" in NV diamond are so far limited to crystals on the order of 5 nm,[20] the presence of this phenomenon and its dependence of the pulse energy would explain our observations, as it could cause the average luminescence intensity to decrease at high pulse energies. It is also possible that the effect is caused by an increased rate of the nonradiative transitions from the electronically excited state.

To explore these possibilities, luminescence decay at different excitation pulse energies was measured for the same crystal. The decay curves for the NV⁻ center are displayed in Fig. 5, panels A and B. As can be seen, the decay rate increases at higher pulse energy (A is low energy, B is high). Quantitatively, the decay does not follow a simple single-exponent law and it takes two exponents as in Eq. (2) to satisfactorily fit the data.

$$R = A_1 \exp(-t/\tau_1) + A_2 \exp(-t/\tau_2), \tag{2}$$

Notably, the parameter most significantly dependent on the energy of the pulse is the ratio of the two amplitudes. The decay times $\tau_1 \approx 7.3$ ns, $\tau_2 \approx 29$ ns and the sum $A_1 + A_2$ do not depend on the pulse energy within the accuracy of the measurements. Although the multi exponential decay could be explained by the fact that the crystal contains many NV-centers, this can not explain the dependence of the decay curve on the laser-pulse energy. The experimental results and the fitting suggest that the emission originates from two states of the NV-center, each with a different decay rate and with energy-pulse dependent relative populations. In particular, the ratio $A_1/A_2$ is 2.4±0.5 for the energy of 1.1 µJ, 0.9±0.3 for 0.060 µJ, and is 0.5±0.2 for the pulse energy of 0.010 µJ (the decay curve is not shown).

Panels C and D of Fig. 5 depict luminescence decay curves taken for NV⁰ centers in the same crystal. Interestingly, we see that in the neutrally charged center the decay fits a single-exponential curve, while maintaining the lifetime shortening seen in NV⁻. However it is possible that the decay curve of NV⁰ follows a two-exponential decay but that it is unobserved due to a smaller difference between the two decay times. But the results shown in Fig. 5 confirm that the decrease of the luminescence intensity is also associated with an increase in the decay rate.

To investigate luminescence recovery, we used modulated excitation pulses to determine the recovery time-scale. The pulse sequence had a period of 10 µs (see Fig. 6, panel A). The energy value $E_m$ was chosen to be close to the energy at the peak of the saturation curve in Fig. 2. Typically this energy was 3 to 5 times the saturation energy. The sequence started with a relatively strong "dimming" pulse, the energy of which varied between $E_m$ and $10E_m$. After this pulses there were *N* "luminescence-recovering" pulses. The energies of these pulses were always $E_m$. Note that this energy is not sufficient to cause a significant luminescence drop but at the same time is large enough to pump practically 100% of the ground state population to the electronically excited state. This simplifies the interpretation of the experimental results.

The dependence of the average NV-luminescence intensity on the energy of the first pulse and the number of the recovering pulses is shown in Fig. 6, panels C and D. The sequence without any "luminescence-recovering" pulses was taken as a reference and all other curves were scaled vertically to overlap at the peak emission rate $R_m$. Such scaling takes into account the trivial increase in photo luminescence rate that results when more excitation pulses are shot within the 10-μs time intervals. As can be seen from Fig. 6, the luminescence drop decreases as the number of recovering pulses increases. To quantify the recovery, we compared the relative decrease of the luminescence rate at different energies of the first pulse. Assuming a simple model with exponential recovery of the luminescence one can write the luminescence rate as

$$R = \frac{R_m}{N+1}\sum_{n=0}^{N}\left(1 - g(E)e^{-bn}\right) = R_m\left(1 - \frac{g(E)}{N+1}\frac{1-e^{-b(N+1)}}{1-e^{-b}}\right), \qquad (3)$$

where g($E$) is the relative drop of the luminescence rate for different values of the pulse energy when the pulse sequence does not include any recovering pulses ($N$ = 0) and $b$ is the recovery constant. Because the separation between the pulses is much larger than the characteristic luminescence time (see Fig. 6), each term in the sum above represents the $n$-th pulse contribution to the total luminescence signal. The term $n = 0$ represents NV-emission collected when only the dimming pulse was present. We fitted the data set for $N$ = 0 pulse sequence with a second order polynomial to get a smooth curve and then calculated the value of g($E$). Thus no particular model of the luminescence diming was selected and the polynomial is simply a guideline. Then Eq. (3) and the curve g($E$) was used to find the emission rate at different values of $N$. The characteristic recovery constant $b$ was the only fitting parameter. This parameter was found to be approximately 0.33 for both sets of data (C and D).

So far the experiment does not distinguish between two distinct processes. The recovery of luminescence could take place independently of the presence of the luminescence-recovering pulses. In such a case, $n$ would be a measure of time passed after "dimming" in microseconds, 1μs being the time interval between pulses. In contrast, light-induced luminescence recovery requires the presence of the pulses and thus in this case *n* counts the recovering pulses irrespective of the time between them and the dimming pulse. To distinguish between these two possibilities, we also measured the luminescence signal with three and four recovering pulses sent 7 µs after the dimming pulse (see Fig. 6, panel B). The corresponding curves in Figure 6 overlap very well with those measured with 1-µs delay and thus confirm that the time delay between the dimming pulse and the recovering pulses was an irrelevant factor.

The effect of the external magnetic field on the emission of the NV-centers was also investigated. The results are shown in Fig. 6. It is clear that the field reduces the magnitude of the intensity drop.

**V. DISCUSSION**

The experiments described above have shown that high-intensity pulsed photo-excitation creates a state of $NV^-$, which has a spectrum characteristic for the photo luminescence of $NV^-$ but a shorter luminescence lifetime. The short-lived state is relatively stable in darkness but can be quickly transformed back to the long-lived state if the center is illuminated with about 3 laser pulses of moderate pulse energy (3-5 times the saturation energy). The creation of the short-lived state seems to have a nonlinear dependence on the laser pulse energy and is not efficient below the energies of about 5 times the saturation energy. Photo ionization of $NV^-$ reported is driven by linear absorption[35] and apparently does not show any significant effect on the relative concentration of the neutral and negatively charged NV centers in our experiments. An external magnetic field reduces the magnitude of the intensity drop. After this brief summary of the main observations, we turn our attention to possible microscopic mechanisms of these phenomena.

It has been recently shown that hydrogen terminated surface of the diamond dramatically reduces the brightness of both types of NV centers while oxidation makes them brighter.[37] Electron transfer to the layer absorbed in the surface creates band bending and conductivity near the surface. The effects of surface termination spreads up to 15 nm in the crystal depth[37] and thus any surface modification could have an effect on the entire crystal which is only about 30 nm in diameter. Surface modification could be responsible for the recently reported decrease of the luminescence at elevated temperatures.[17] If a specific termination (needed to explain the results) is formed at high levels of exposure, is stable in darkness for much longer than 10 µs, can be reverted with a very few moderate energy pulses of light as shown in Fig. 6, there mush be at least two distinct mechanisms involved in the process (one being activated only at high pulse energies). None of these mechanisms was reported in the literature. Therefore we propose a different mechanism to explain such a dual role of light, which is much more based on the known effects and also explains the effect of the applied magnetic field.

The longer decay time of 29 ns is close to the values measured for the radiative decay rate of NV-centers in nano-crystals.[38] We attribute this rate to the radiative decay of m=0 spin state, and the 7-ns exponent to the mostly non-radiative decay of the ±1 spin states. Because $A_1/A_2 = 0.5$ at low laser-pulse energy, we conclude that the population of the highly luminescent *m* = 0 spin state is about 70% and the total population of the other two spin states is 30%. The 70% spin polarization is close to the values reported in the literature.[39] As the energy increases significantly beyond the saturation value, the population of the $m=\pm 1$ states rises and the spin state gets spin depolarized. Ultimately equal populations of all three sublevels are achieved. This explains the leveling of the saturation curve at high pulse energies. When the short-lived exponent has its relative amplitude close to 2/3 (1/3 for $m=1$ and 1/3 for $m=-1$) no further reduction in the emission is possible. Note that 2/3 is very close to 0.7, the relative amplitude of the short-lived exponent at high pulse energy, as observed in the experiment $(1/(1 + A_2/A_1) \approx 0.7)$. The recovery of the luminescence as shown in Figure 6 is explained by light induced spin repolarization which restores the preferable population of $m = 0$ state. The 1.8-fold decrease in the luminescence signal indicates in our model that the non-radiative relaxation rate to the singlet state from the $m = \pm 1$ spin-sublevels is approximately 2.5 times higher than the photon emission rate. Therefore if the population of the singlet state relaxed entirely to the $m = 0$ spin sublevel of the electronic ground state as conventionally assumed,[25, 40] then it would take one polarizing pulse to reduce the population of $m = \pm 1$ spin-sublevels by a factor of 3.5. The observed recovery constant of 0.3 indicates that it takes 3 spin-polarizing pulses to reduce the depolarization effect by a factor of 2.7. This, much slower spin polarization has been recently observed in an independent experiment[41] and can be explained by assuming that the relaxation from the singlet can proceed to any of the three spin sublevels of the ground state with equal probabilities in line with some theoretical predictions.[29]

The last of the described experiments unambiguously demonstrates that when the spin polarization is essentially reduced by the applied external magnetic field, the drop in the emission at high pulse-energies significantly decreases. The reduction of the spin polarization in the presence of the magnetic field results from mixing the states with different values of the spin projection on the axis of the NV center.[42] Note that the saturation curve measured when the field was switched on is much closer to the prediction of Eq. (1) but that the luminescence intensity at the highest level of the pulse energy is close to the intensity observed without the field. This is expected because the right end of the saturation curve corresponds to small spin polarization in any case. Interestingly, the change in the spine polarization does not affect the saturation energy value. This insensitivity of the saturation curve to the population of the metastable states is expected for the short-pulse excitation.[32]

Spin-lattice relaxation is a known mechanism of spin depolarization. The spin-lattice relaxation time at room temperature is about 1 ms but above room temperatures it quickly shortens, proportionally to the fifth power of the temperature.[43] Because the zero-phonon line changes insignificantly under high pulse energy excitation, a high temperature may last only a small fraction of the luminescence lifetime. Achieving the thermal equilibrium of the spin states within a few nanoseconds requires unrealistically high temperatures (10 times higher than the room temperature). But it is interesting to discuss the observed results in relation to the photo-ionization of the negatively charged NV centers.[35] First, the absence of significant photo induced changes in the relative populations of the two species, $NV^-$ and $NV^0$, is puzzling. The second question is how the photo ionization process interferes with the spin polarization mechanism.

Notably, the photo ionization efficiency of $10^{-3}$ reported[35] is too small to have any significant effect on the spin polarization, which requires only several absorption-relaxation cycles for completion. However, excitation from the first excited triplet state (note that the drop in the luminescence intensity starts well above the saturation energy), promotes the optical electron into the conduction band of the host crystal which is less than 1.5 eV above the excited triplet state of the center[29] while the energy of the 532-nm photon is 2.3 eV. This should make the ionization much more efficient. Thus, one could expect $NV^- \rightarrow NV^0$ conversion efficiency close to 1. Given that the inverse process $NV^0 \rightarrow NV^-$ has a characteristic lifetime of 10 μs[35] and is much longer than the time between the laser pulses, the decrease of the number of $NV^-$ species and increase of the concentration of their neutral counterpart should be very strong and significant but is not observed. A natural explanation for this puzzle comes from the relatively high concentration of NV centers in these crystals. A process in which an electron is stripped from the negative center and transferred to the neutral center does not actually change the number of centers of each kind but dynamically converts one type into another. This presents a plausible path by which the $NV^-$ centers can be observed to lose their spin polarization. Recent papers[44, 45] studying the $NV^0$ center have found that it has a relatively similar electronic structure, albeit with the ground state and the optically excited state being quadruplets. It has been experimentally verified[45] that the neutral NV center can also be spin polarized. This similarity lends credence to the idea that photoionization is involved in the observed luminescence drop of both centers. Although spin depolarization is not likely to be the only mechanism behind the luminescence drop, the proposed mechanism must have a significant role.

**VI. CONCLUSION**

In conclusion, we have shown that our observations of the saturation curves, luminescence decay rates, the response to the modulated pulse sequence, and the effect of the magnetic field on the saturation curves (including the part at high energies of the excitation laser-pulses) can be explained by assuming optical polarization and depolarization of the electron spin in the $NV^-$ centers and relaxation from the singlet state to $m = 0, \pm 1$ spin states of the ground triplet with approximately equal probabilities. Excitation to a higher electronic state from the first excited triplet state (in particular to the conduction band of the host crystal) followed by a spin depolarizing nonradiative return back to the first excited triplet state is a possible pathway for the depolarization process. We suggest that this process is facilitated by transferring the electron from $NV^-$ in to a nearby $NV^0$ center, the process preserving the total number of centers by transporting the electron form between two locations. The presented results point to a new direction of the research on NV-centers in nano-diamond focused on inter-center coupling effects and spin transport properties. The current work points to exciting perspectives in the spintronics of the NV-centers in diamond. For example, one can think of transportation of an electron through a closely spaced and purpose-specifically engineered network of NV centers.

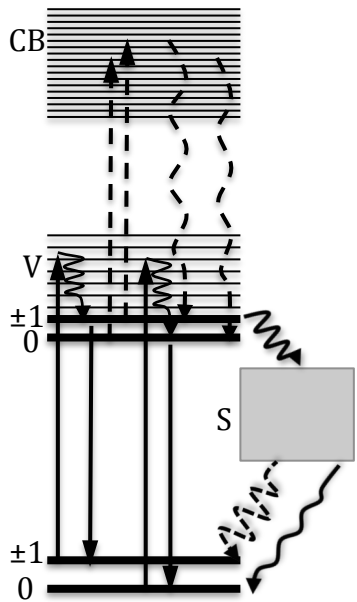

**Figure 1**. Simplified energy diagram of an NV-center. The straight lines represent radiative transitions (absorption and emission of a photon). Wavy lines symbolize non-radiative transitions. Solid lines sketch the conventional path for transferring population from $m = \pm 1$ states to $m = 0$ states and dashed lines represent transition proposed in this paper to explain our experimental observations (they transfer a part of the population from m = 0 back to m = ±1). All the singlet states are shown as a gray box and labeled S. CB stands for conduction band and the crystal phonons contribute to the states denoted by V.

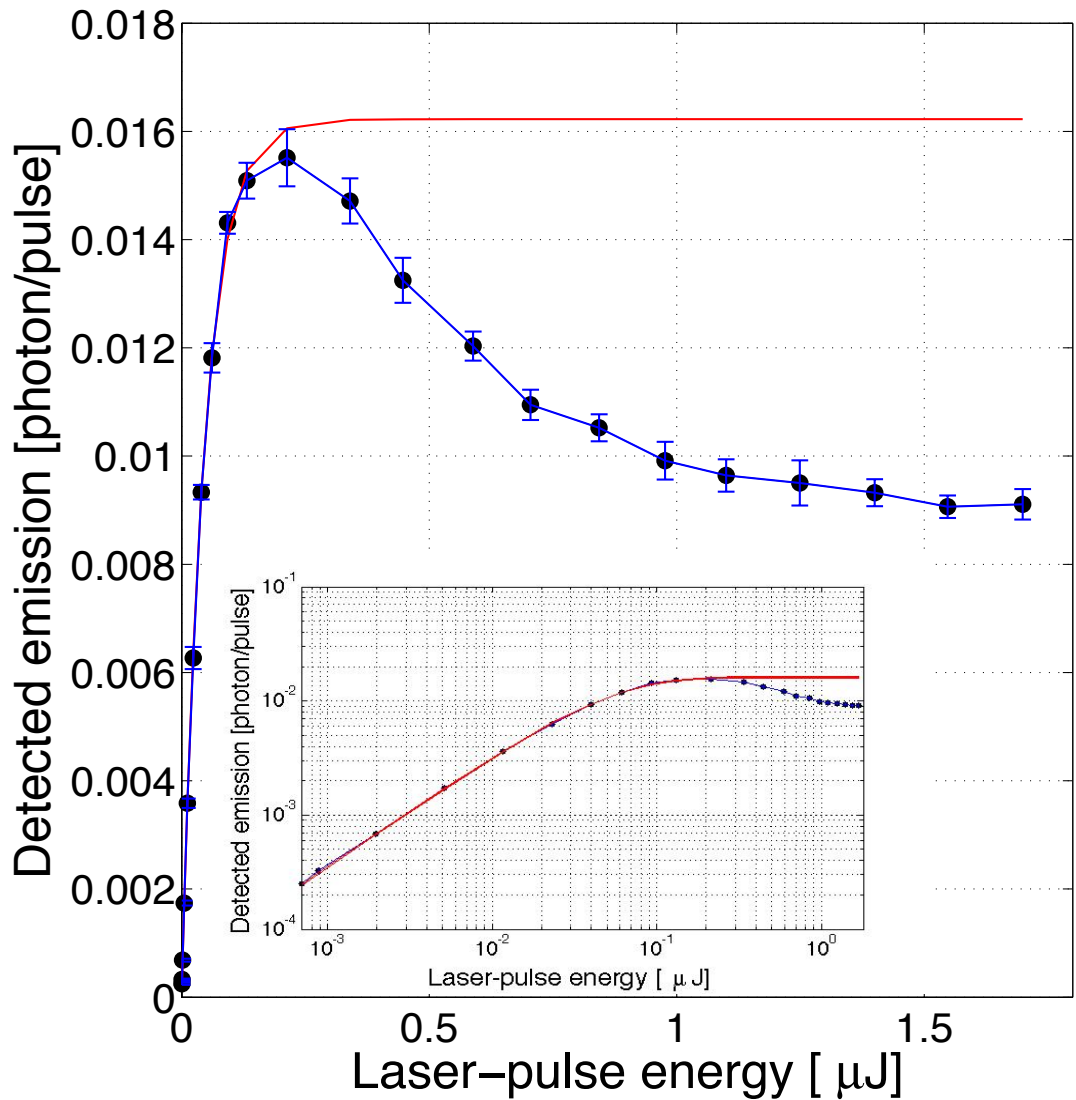

**Figure 2.** Photo luminesces of NV⁻ centers at different energies of the exciting laser pulse (pulse repetition rate is 1 MHz). The insert represents the same data in a log-log plot for better visibility of the low-energy part of the curve. The smooth curve is obtained by fitting Eq. (1) to the first 8 low-energy points.

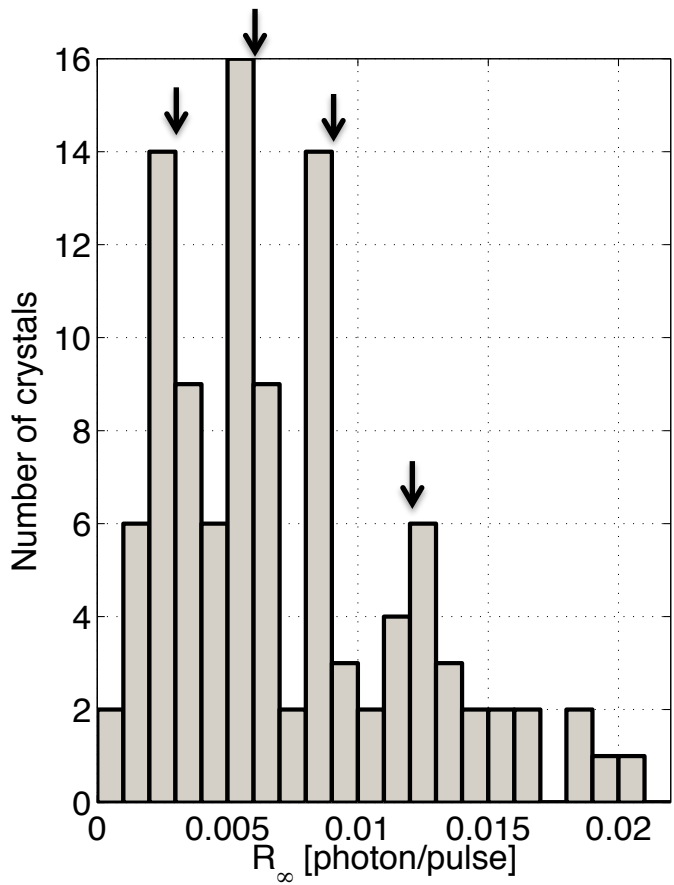

**Figure 3** A histogram of the maximum emission rates for the batch of the crystals used in the experiments. Four distinct peaks (indicated by arrows separated horizontally by 0.003 photon/pulse) correspond to 1, 2, 3, and 4 negatively charged NV-centers in the crystals.

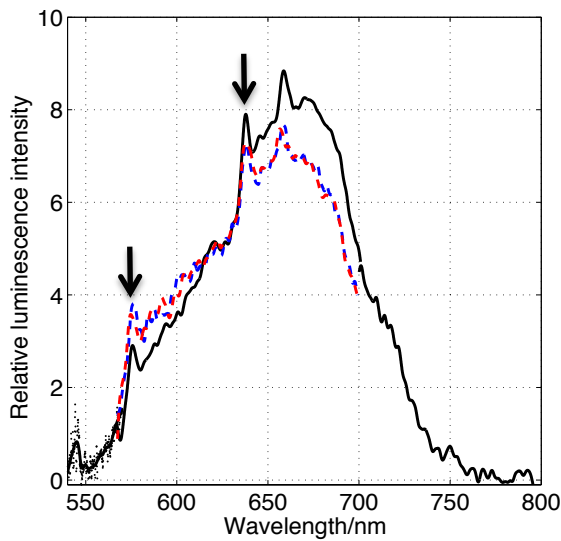

**Figure 4.** Luminescence spectra of NV-centers in the same nano crystal as shown in Fig. 2 measured at different excitation energies: 0.13 μJ (solid line), 0.34 μJ, and 0.8 μJ (both are shown as dashed lines). Small distortion (about 15% rise on the left hand side and corresponding 15% fall on the right hand side) may be attributed to a shift in the dynamic balance between neutral and negative forms of the NV-centers but can not explain the two-fold decrease of the $NV^-$ emission as shown in Fig. 2. Arrows indicate zero phonon lines – NV⁰ (~580 nm) and NV⁻ (~638 nm). A drop in one zero phonon line without a corresponding decrease in the other would imply population transfer between the two species.

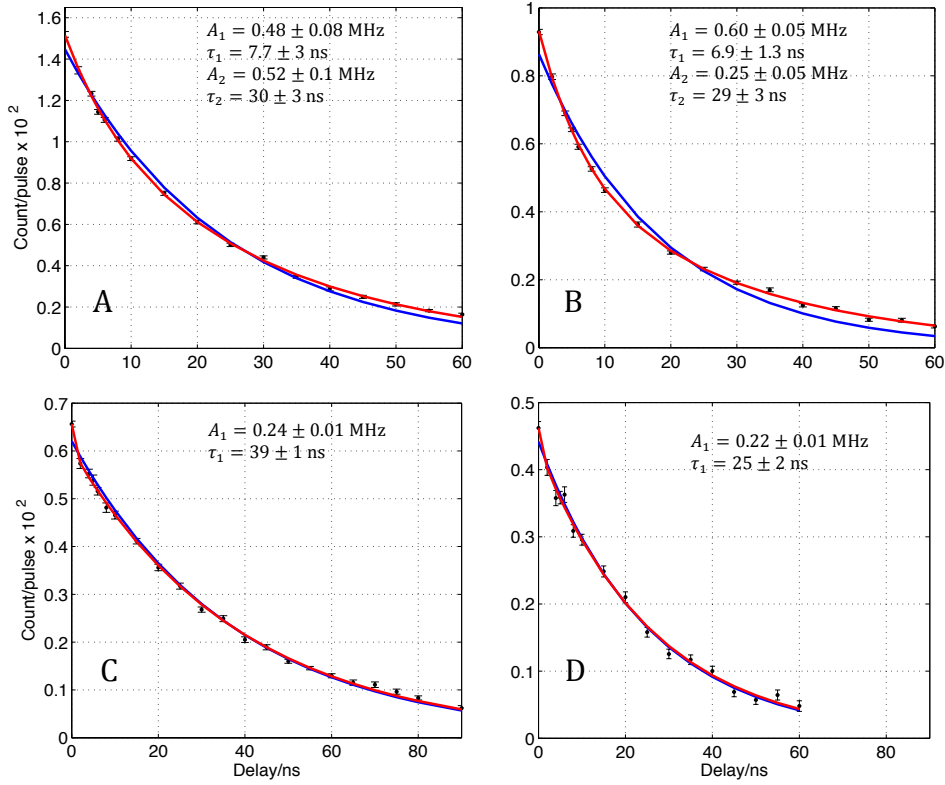

**Figure 5.** Decay curves measured at different energies of the exciting laser pulse for the emission at 600 nm and 700 nm. Panels A (pulse energy 0.13 µJ ) and B (pulse energy 1.1 µJ ) show the data and the fitted curves for 700-nm emission. Panels C (pulse energy 0.060 µJ) and D (pulse energy 1.1 µJ) characterize 600-nm emission. The fit was done using one and two-exponential decay models. The signal was integrated over a time interval of $\Delta t = 40$ ns. Therefore the number of counts per pulse reads $A_1\tau_1\left[1-\exp(-\Delta t/\tau_1)\right]\exp(-t/\tau_1) + A_2\tau_2\left[1-\exp(-\Delta t/\tau_2)\right]\exp(-t/\tau_2)$, where $t$ is the delay. The best fits to one-exponential decay in panels A and B result in $\chi^2 = 10$ and $\chi^2 = 26$ respectively ($\chi^2 \approx 1$ for the two-exponential fits in both cases). Data shown in Panels C and D can be satisfactorily fitted to one-exponential curves. The parameters of the fits and their 95% confidence intervals are shown in the panels.

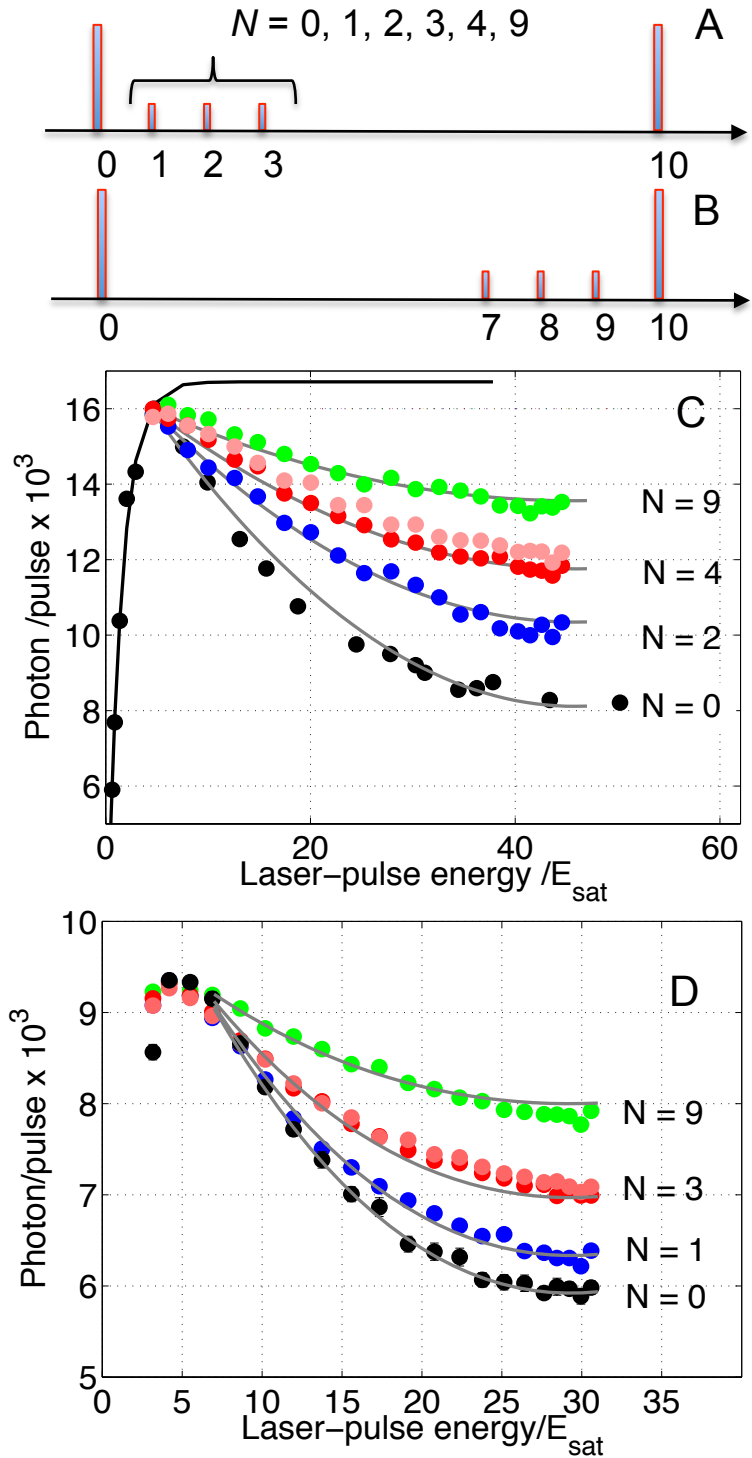

**Figure 6** Panel A shows a pulse sequence where 3 (N = 3) relatively weak, spin-polarizing pulses separated by a 1-μs delay follow a strong "depolarizing" pulse. In panel B, the sequence of the 3 pulses comes after a time delay of 7 μs. Panels C and D show how the average luminescence intensity depends on the number of weak pulses and the energy of the first pulse in the sequence. Curves with N = 4 and 3 shown in panels C and D respectively were first measured with quadruplets/triples of pulses coming as shown in panel A and then as shown in panel B. Small variation between the two corresponding curves indicates that the waiting time was not an essential parameter. The recovery of the signal depends on the number of the weak pulses not on the time passed after the strong pulse. The gray solid curves in panels C and D are obtained using Eq. (3) as explained in the text and show that the recovery is light driven with exponential recovery rates of 0.38 and 0.28 per pulse respectively.

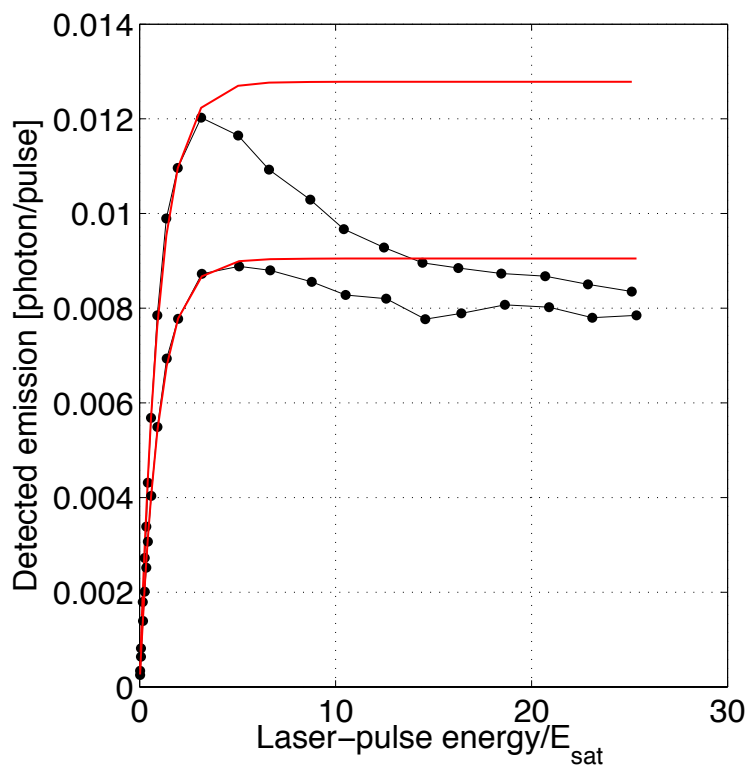

**Figure 7.** Two saturation curves measured with switched on and off external magnetic field. The upper set of data points corresponds to the "external field off" situation. The lower set was measured when a 1000-G magnetic field was present. The direction of the field is parallel to the optical axis of the microscope objective collecting the luminescence. The two solid lines are the fits of Eq. (1) to the data points corresponding to 11 ("field off" set) and 12 ("field on" set) lowest values of the laser-pulse energy.